\documentclass[journal=jpcafh,manuscript=article,layout=twocolumn]{achemso}

\usepackage[version=3]{mhchem} 
\usepackage{braket}
\usepackage[utf8]{inputenc}

\newcommand{\pp}{$\pi \pi ^{\ast}$}
\newcommand{\ps}{$\pi \sigma ^{\ast}$}

\author{Daniel Kinzel}
\author{Philipp Marquetand}
\email{philipp.marquetand@univie.ac.at}
\author{Leticia Gonz{\'a}lez}
\affiliation[Uni Jena]
{Institut f\"ur Physikalische Chemie, Friedrich-Schiller-Universit\"at Jena, 
Helmholtzweg 4, 07743 Jena, Germany.}

\title[\texttt{achemso} demonstration]
{Stark control of a chiral fluoroethylene derivative}

\begin{document}
\begin{abstract}
Hydrogen dissociation is an unwanted competing pathway if a torsional motion 
around the C=C double bond in a chiral fluoroethylene derivative, namely (4-methylcyclohexylidene) fluoromethane (4MCF), is to be
achieved. We show that the excited state H-dissociation can be drastically
diminished on timescales long enough to initiate a torsion around the C=C double
bond using the non-resonant dynamic Stark effect. Potential energy curves,
dipoles and polarizabilities for the regarded one-dimensional reaction
coordinate are calculated within the CASSCF method. The influence of the
excitation and the laser control field is then simulated using wavepacket
dynamics.
\end{abstract}

\section{Introduction}
Laser control of chemical reactions has been on the cutting edge of current
research for several years already but is still in
its beginnings~\cite{Brumer1992ARPC,Gordon1997ARPC,Rice1997ACP,Tannor1999FD,
Rice2000,Dantus2001ARPC,Brixner2001AAMOP,Shapiro2003RPP,Shapiro2003,Daniel2003SCI,Dantus2004CR,Hertel2006RPP,
Nuernberger2007PCCP,Engel2009ACP,Worth2010PCCP}. One of the control strategies is to make use
of the Stark effect~\cite{Stark1913NAT} where the molecular potentials are
considerably distorted to yield dressed states or light-induced potentials
(LIPS)~\cite{Garraway1998PRL}. With static fields, oriented samples can be
prepared, where the eigenstates are called pendular
states~\cite{Stapelfeldt2003RMP}. Interesting rovibrational dynamics can then be
observed after photoexcitation, see e.g.
Refs.~\cite{Marquetand2004JCP,Marquetand2005PCCP}. The Stark effect also plays a
role in the interaction with oscillating electric fields produced by lasers. If
the laser frequency is high enough, the states mainly follow the field envelope.
This shift is known as the dynamic Stark effect, which is particularly
interesting for quantum control since it works also when the laser is
non-resonant~\cite{Sussman2006SCI}. Especially the latter case is interesting
because no highly specific wavelength sources are required, and this non-resonant
dynamic Stark effect (NRDSE) has already been the target of several
studies~\cite{Levis2001SCI, Sussman2006SCI, Gonzalez-Vazquez2006CPL,
Gonzalez-Vazquez2006JPCA, Chang2009JMO, Chang2009JCP_2, Gonzalez-Vazquez2009JCP,
Gonzalez-Vazquez2010PCCP, Sussman2011AJP, Townsend2011JPCA}.
There, reaction pathways are reversibly changed and in this way, the NRDSE acts 
like a photonic catalyst.

A prominent control target is the \textit{cis}/\textit{trans} photoisomerization
of an olefinic double bond since in this way chemical properties determined by E-Z isomerism can be changed or energy can be transformed into molecular
motion~\cite{Levine2007ARPC}. The latter effect is the basis for molecular
engineering in nanotechnology, where molecular switches, rotors and motors are
investigated~\cite{Balzani2009CSR}.

\begin{figure}
 \includegraphics[width=6cm,keepaspectratio]{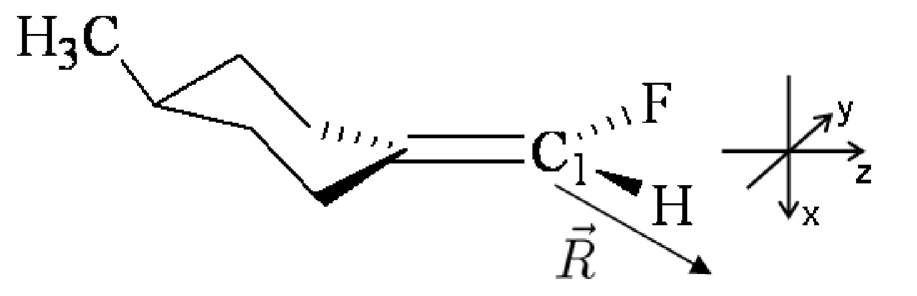}
 \caption{\label{fig:4MCF} \footnotesize{Structural formula of
(4-methylcyclohexylidene) fluoromethane; $\vec{R}$ is indicating the
dissociation vector.}}
\end{figure}

While several studies use simplified models to get insight into the dynamics and
control of light driven
rotors, see e.g. Refs.~\cite{Hoki2003ACIE,Hoki2004JPCB, Hoki2004JPCA,Fujimura2004CPL, Marquetand2006JCP,Yamaki2009PCCP,Perez-Hernandez2010NJP}, we want
to point out another aspect which is important in this context: Besides the
turning of the rotor, competing processes can play a role but should be avoided. In our case, the
turning motion is the rotation around the double bond of the chiral fluoroethylene
derivative
(4-methylcyclohexylidene) fluoromethane (4MCF),
see \ref{fig:4MCF}.
In a series of papers, we have introduced 4MCF as a molecular rotor/switch and
investigated different adversary pathways to the desired turning which consists of
a switching between the R/S
enantiomers~\cite{Kroener2003PCCP,Kroener2004CP,Fujimura2004CPL,Alfalah2010CP,
Kinzel2011IJQC,Kinzel2011TBP}. Interestingly, a recently discovered conical
intersection (CI) between a $\pi\pi^\ast$ and $\pi\sigma^\ast$ state at the
Franck-Condon (FC) geometry allows for different dissociation
channels in the electronically excited state~\cite{Kinzel2011IJQC}. Our subsequent results from semiclassical
simulations in full dimensionality indicate that the ultrafast H-dissociation is the most
important reaction channel after laser excitation~\cite{Kinzel2011TBP}.
Therefore, in this paper, we will show how Stark control can be employed to
prevent the molecule from being destroyed.
%


\section{Computational Details}

We consider a one-dimensional (1D) model for the dissociation of the hydrogen belonging to the fluoromethane-moiety (\ref{fig:4MCF}). Our previous semiclassical full-dimensional
simulations \cite{Kinzel2011TBP} show that the dissociation of this hydrogen proceeds along its bond axis.
The reaction happens very fast, and as such, the rest of
the molecular framework keeps nearly unchanged during the dissociation process. Hence, a 1D approach is justified here, with the reaction coordinate pointing along the C$_1$-H axis.

%
%
%
The potential energy curves (PECs) for this coordinate are calculated using the state-averaged complete active space
self-consistent field method (SA-CASSCF) \cite{CASSCF_Roos_ACP87}. The active
space employed contains 4 electrons in 4 orbitals, namely the $\pi_{CC}$, the
$\sigma_{CH}$ and their corresponding antibonding ones. This active space is sufficient to calculate
the three lowest lying singlet electronic states which correspond to the ground state, the spectroscopic bright {\pp} state and the {\ps} state playing a major role in the dissociation process. The basis set
used is the double-zeta polarized Pople basis set 6-31G*~\cite{HehreWILEY1986}.
This rather small basis set is used intentionally in order to avoid Rydberg
mixing (see Ref. \cite{Schreiber2007JPCA}).

When constructing the PECs, the molecule is assumed to be preoriented such that
the laboratory z-axis lies within the C=C double bond. The fluorine and hydrogen
atom connected to the C$_1$ of this double bond are then found in the yz-plane. Within this
preorientation, the dissociation coordinate, further on labeled $R$, is defined as
the distance between the double bonded carbon $C_1$ and the attached hydrogen atom,
see \ref{fig:4MCF}. The rest of the molecular framework is kept frozen at
the geometry optimized at the MP2/6-311+G(d,p) level of theory as described in
Ref. \cite{Schreiber2007JPCA}. A grid of a total of 18 points is calculated along
$R$ using the SA3-CASSCF(4,4) / 6-31G* protocol as implemented in the MOLPRO
program package \cite{MOLPRO}. The grid points are equally distributed between
$R=$ 0.6~{\AA} and $R=$ 3.0~{\AA} with an spacing of 0.2~{\AA}. Additional points
were added to take account for the crossing around the Frank-Condon
point at 1.05, 1.08, 1.10 and 1.15~{\AA}. To ensure a correct asymptotic behaviour
of the resulting PECs an extra grid point at $R=$ 50~{\AA} was added. 
Furthermore, permanent dipoles $\mu_{i}$, transition dipole moments between 
each state $\mu_{ij}$, and polarizabilities $\alpha_{ij}$ have been computed.
Here, the polarizabilities have been evaluated numerically according to the
MOLPRO manual~\cite{MOLPRO}. All potentials and corresponding properties were
cubic-splined to give 1024 points between $R=$ 0.6~{\AA} and $R=$ 3.0~{\AA}.

Since at least one CI plays a role during the
deactivation process, non-adiabatic coupling terms (NACTs) $T_{ij}^R$ between
each state, $i$ and $j$, with respect to $R$, defined as
\begin{equation}
 \label{eqn:nacts}
 T^{R}_{ij}=\langle \chi_i|\frac{\partial}{\partial R}\chi_j\rangle,
\end{equation}
were calculated at the same level of theory using a three-point formula as
implemented in MOLPRO~\cite{MOLPRO}.\\


In order to investigate the dynamics of the dissociation process in 4MCF, we
solve the time-dependent Schr\"odinger equation (TDSE) for the nuclei in each of
the three states. In the adiabatic representation, the TDSE for the three-state model is written as 
\begin{multline}
 \label{eqn:TDSEa}
i \hbar \frac{\partial}{\partial t} \left( \begin{array}{c}
\psi^{ad}_0(t) \\ \psi^{ad}_1(t) \\ \psi^{ad}_2(t) \end{array} \right) =\\
\left(\begin{array}{ccc} {H}^{ad}_{00} & {H}^{ad}_{01} & {H}^{ad}_{02} \\
{H}^{ad}_{10} & {H}^{ad}_{11} & {H}^{ad}_{12} \\
{H}^{ad}_{20} & {H}^{ad}_{21} & {H}^{ad}_{22} \end{array} \right) \left(
\begin{array}{c} \psi^{ad}_0(t) \\ \psi^{ad}_1(t) \\ \psi^{ad}_2(t) \end{array}
\right).
\end{multline}
with approximate matrix elements of the Hamiltonian given as
\begin{equation}\label{Had}
\begin{split}
&{H}^{ad}_{ii}=-\frac{\hbar^2}{2M} \frac{\partial^2}{\partial R^2} + V_i^{ad}
~~~~~ \mathrm{and}\\
&{H}^{ad}_{ij}=-\frac{\hbar^2}{M} ~ T^{R}_{ij}\frac{\partial}{\partial R},
\end{split}
\end{equation}
in which the kinetic couplings are expressed as in \ref{eqn:nacts}. $V_i^{ad}$ are the
electronic adiabatic states computed as described above and $M$ is the reduced
mass between the hydrogen atom and the rest of the molecule. The second order
kinetic couplings defined as
\begin{equation}
 \label{eqn:nacts2}
 T^{RR}_{ij}=\langle \chi_i|\frac{\partial^2}{\partial R^2}\chi_j\rangle,
\end{equation}
are neglected since they are much smaller than the first order ones.

Describing adiabatic nuclear dynamics in the presence of CIs
is a difficult task, since the NACTs at these points eventually become
singularities. Treating such sudden changes in the character of the
wavefunction numerically is very challenging. To overcome this problem nuclear
dynamics will be carried out in the diabatic representation. Here, we use a
unitary transformation matrix $\mathbf{U}$ to derive the diabatic potentials
from the
adiabatic ones:
\begin{equation}
 \label{eqn:UdagPU}
 \mathbf{V}^d=\mathbf{U}^{\dagger} ~ \mathbf{V}^{ad} ~ \mathbf{U}
\end{equation}
The same applies for the transformation from the adiabatic to the diabatic
dipole as well as polarizability matrix, $\mathbf{\mu}^{ad/d}$ and 
$\mathbf{\alpha}^{ad/d}$, respectively.

The coordinate-dependent transformation matrix $\mathbf{U}(R)$ has been numerically derived by using the
Crank-Nicholson-like equation for the transformation matrix propagation as
described in Ref. \cite{Esry2003PRA}, which is written as
\begin{multline}
 \label{eqn:cranknicholson}
 \left(\mathbf{I}+\mathbf{T}\frac{\Delta R}{2}\right) ~ \mathbf{U} (R+\Delta R)
 =\\
 \left(\mathbf{I}-\mathbf{T}\frac{\Delta R}{2}\right) ~ \mathbf{U} (R),
\end{multline}
where $\mathbf{I}$ is the identity matrix and $\mathbf{T}$ is the matrix
containing the nonadiabtic coupling elements $T_{ij}^R$, see \ref{eqn:nacts}.

Here, the matrix $\mathbf{U}$ will be propagated in the spatial coordinate along
the 1D potential starting at $R=$ 3~{\AA} and evolving until $R=$ 0.6~{\AA}.
Thus, at $R=$ 3~{\AA}, the transformation matrix is set to be the identity
matrix, i.e. $\mathbf{U}(R=\mathrm{3}$~{\AA}$)=\mathbf{I}$. As in our case
$\mathbf{U}$ is propagated backwards along $R$ we set $\Delta R = -\Delta R$.

After transformation, $V_i^d$ are then the diabatic potentials for the three
states of interest and $V_{ij}^d=V_{ji}^d$ are the potential couplings. Hence,
in the diabatic representation, the TDSE is then written as
\begin{multline}
 \label{eqn:TDSEd}
i \hbar \frac{\partial}{\partial t} \left( \begin{array}{c}
\psi^{d}_0(t) \\ \psi^{d}_1(t) \\ \psi^{d}_2(t) \end{array} \right) 
=\\
\left(\begin{array}{ccc} {H}^{d}_{00} & {H}^{d}_{01} & {H}^{d}_{02} \\
{H}^{d}_{10} & {H}^{d}_{11} & {H}^{d}_{12} \\
{H}^{d}_{20} & {H}^{d}_{21} & {H}^{d}_{22} \end{array} \right) \left(
\begin{array}{c} \psi^{d}_0(t) \\ \psi^{d}_1(t) \\ \psi^{d}_2(t) \end{array},
\right).
\end{multline}
with
\begin{equation}\label{Hd}
{H}^{d}_{ii}=-\frac{\hbar^2}{2M} \frac{\partial^2}{\partial R^2} + V_i^{d}
~~~~~ \mathrm{and} ~~~~~
{H}^{d}_{ij}=V_{ij}^{d},
\end{equation}

The elements $V_i^{d}$ and $V_{ij}^{d}$ form the diabatic
potential matrix $\mathbf{V}^d$.
In the presence of an external electric field $E(t)$, this
potential matrix is replaced by a matrix $\mathbf{W}$
whose elements are given by
\begin{equation}
 \label{eqn:W}
 W_{ij}=V_{ij}^d - \mu_{ij}^d ~ E(t) - \alpha_{ij}^d ~ E(t)^2.
\end{equation}
Diagonalizing the matrix $\mathbf{W}$ results in the so-called dressed states potentials
$V_i^{dressed}$.\\

In this paper, the total dynamic electric field that affects the molecule is
modeled as a sum of a resonant Gaussian-shaped UV pulse and a non-resonant strong field control pulse of approximately rectangular shape, 
\begin{multline}
 \label{eqn:Efield}
 \vec{E}(t)=\vec{\epsilon}^{~UV} E_0^{UV} G(t) \cos(\omega^{UV} t) \\
 + \vec{\epsilon}^{~control} E_0^{control} S(t) \cos(\omega^{control}t),
\end{multline}
with the polarization vectors $\vec{\epsilon}^{~UV/control}$, the field amplitudes
$E_0^{UV/control}$, the frequencies $\omega^{UV/control}=\frac{2\pi c}{\lambda^{UV/control}}$ with $c$ being the speed of light, the Gaussian envelope function $G(t)$ and the
analytical shape function $S(t)$ defining the envelope of the control pulse.
To mimick a realistic rectangular shape, $S(t)$ is described by a $\sin^2$ type function from the beginning of the
pulse at $t_i$ until a constant value of $S(t)=$ 1 is attained at time $t_{c1}$. The pulse is switched off in the same fashion from time $t_{c2}$ until the end of the pulse $t_f$.
In the case of a static external field \ref{eqn:Efield} reduces to
$\vec{E}(t)=\vec{\epsilon} E_0$.

The diabatic TDSE including the field interaction, \ref{eqn:TDSEd},\ref{eqn:W}, is
solved with the help of the split-operator method
\cite{Feit1982JCOMPP,Feit1983JCP,Feit1984JCP,Kosloff1994ARPC} with a time
discretization of $\Delta t=$0.01 fs.

The system is initially prepared in the vibrational ground state of $V_0^{ad}$ computed with the Fourier-Grid-Hamiltonian method (FGH)~\cite{Martson1989JCP}.

In order to prevent artificial reflections of the wavepacket from the grid
boundary, a cut-off function, $\gamma(R)$, is introduced, which
annihilates parts of the outgoing wavefunction $\psi_i$ in each state at the
end of the grid. This function is defined as
\begin{multline}
 \label{eqn:cutoff}
 \gamma(R)=\\
 \biggl\lbrace \begin{array}{cc} cos^2\left[\frac{\pi}{2}
\frac{R_{end}-R_{\gamma}-R}{R_{\gamma}}\right] & \mathrm{if} ~~ R > R_{end}-R_{\gamma} \\ 1 &
\mathrm{otherwise} \end{array},
\end{multline}
with $R_{end}=$ 3~{\AA} being the end of the grid, and the cut-off parameter
$R_{\gamma}$ which we set to 0.5~{\AA}, meaning that the cut-off function starts at
$R=$ 2.5~{\AA}.

On the basis of this cut-off function, we define the accumulated flux for each
state, $I_i^{acc} (t)$, as the part of the wavefunction that has been cut off
after the previous time step,
\begin{equation}
 \label{eqn:flux}
 I_i^{acc} (t)=\sum_t \left|\psi_i (t-\Delta t)\right|^2 -
 \left|\psi_i (t)\right|^2.
\end{equation}

\section{Results and Discussion} 
\label{sec:results}

\subsection{Field free potential energy curves}

\begin{figure}
\centering
\includegraphics[width=8cm,keepaspectratio]{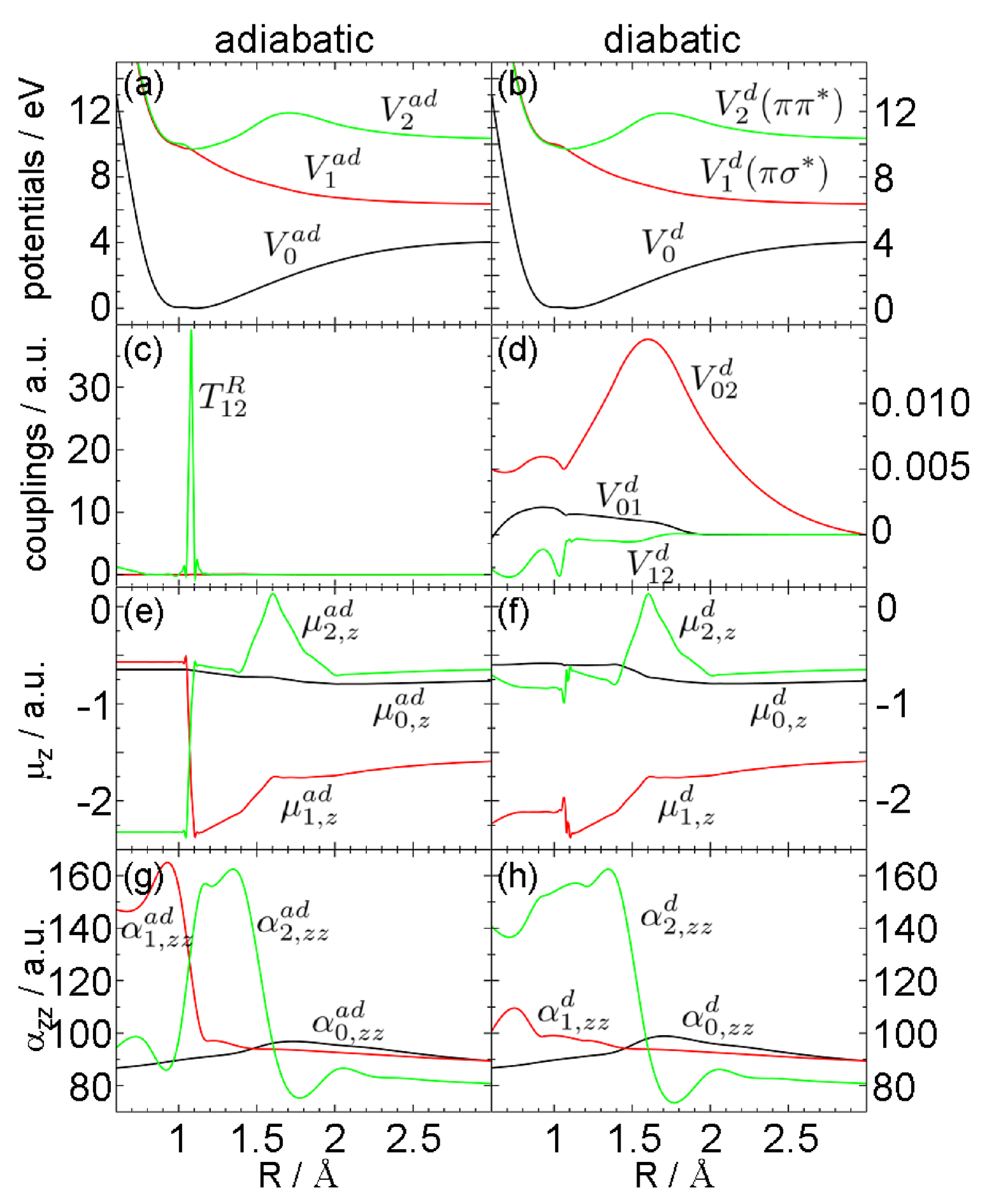}
\caption{\label{fig:1D} \footnotesize{Adiabatic (a) and diabatic
(b) potential energy curves, corresponding kinetic (c) and potential (d)
couplings, z-polarized adiabatic (e) and diabatic (f) permanent dipole moments
and zz-polarized adiabatic (g) and diabatic (h) polarizabilities}}
\end{figure}

In \ref{fig:1D}, we present the one-dimensional adiabatic and
diabatic potential energy curves (PECs), $V_i^{ad}$ and $V_i^{ad}$, their corresponding
kinetic and potential couplings, $T_{ij}$ and $V_{ij}$, respectively, as well as
the adiabatic and diabatic z-polarized permanent dipole moments and zz-polarized
polarizabilities, $\mu_i^{ad/d}$ and $\alpha_i^{ad/d}$, respectively.

The adiabatic potential (\ref{fig:1D}a) shows a near degeneracy point
between states $V_1^{ad}$ and $V_2^{ad}$ at the Frank-Condon distance of $R=$
1.08~{\AA} which coincides nicely with our earlier findings in Ref.
\cite{Kinzel2011TBP}. At that point the nonadiabtic coupling term $T_{12}^R$
amounts to 39.2 a.u., indicating a strong coupling between these states.
Furthermore, the character of the electronic wavefunction of $V_1^{ad}$ switches
from {\pp} to {\ps} and vice versa of that of $V_2^{ad}$. This switch is moreover
illustrated by the adiabatic permanent dipole moments, $\mu_1^{ad}$ and
$\mu_2^{ad}$, and adiabatic polarizabilities, $\alpha_1^{ad}$ and $\alpha_2^{ad}$,
(\ref{fig:1D} e and g, respectively) that switch their values at the point of
degeneracy. These findings coincide with the results from Ref. \cite{Kinzel2011IJQC} stating that the FC geometry is actually a point on the multidimensional seam of CIs between the states $V_1^{d}$ and
$V_2^{d}$.\\ 

\subsection{Dressed states in the presence of an external field}

\begin{figure*}
\centering
\includegraphics[width=12cm,keepaspectratio]{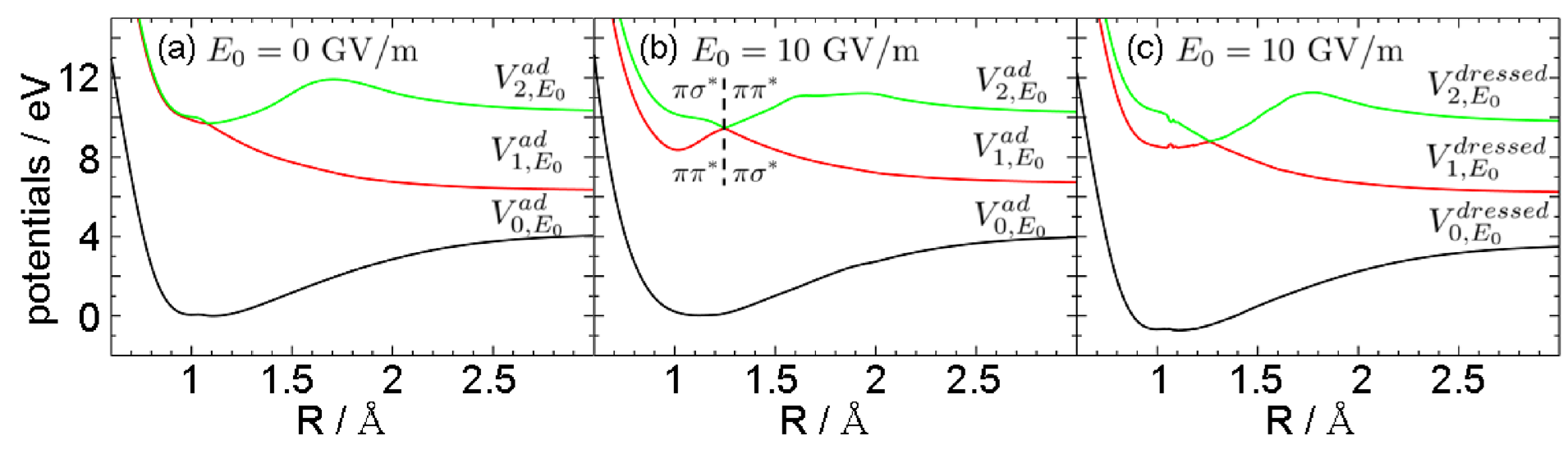}
\caption{\label{fig:1D-field} \footnotesize{(a) Unperturbed adiabatic PECs;
(b) perturbed PECs obtained by adding the field interaction in the electronic
Hamiltonian; (c) dressed PECs obtained by diagonalizing
$V_{ij}^d-\mu_{ij,z}^d E_0-\alpha_{ij,zz}^d E_0^2$}. The change of the electronic character in the adiabatic states is exemplarily indicated in panel (b).}
\end{figure*}

In order to prevent H-dissociation, we want to employ the NRDSE to trap the wavepacket in
the {\pp} state. Since the double bond of the molecule lies in the z-axis,
all fields and field interactions are assumed to be z-polarized in the following. As a
first attempt to study the effect of a strong field on the PECs, a static
electric field with a field strength of $E_0=$ 10~GV/m is added to the electronic
Hamiltonian as implemented in the MOLPRO program package \cite{MOLPRO}. Each
point of the PECs is recalculated and the resulting new
potentials are shown in \ref{fig:1D-field}b. Compared to the unperturbed
potentials (\ref{fig:1D-field}a) the crossing point has been shifted
to longer C-H distances (1.25~{\AA}). This leaves the opportunity to trap a
wavepacket at smaller distances on the first excited state potential $V_{1,E_{0}}^{ad}$ after excitation from the ground state. Note that the electronic character in this region is {\pp} (as exemplarily indicated in \ref{fig:1D-field}b) and therefore, the majority of the overall population will be excited to this bright state.
However, static fields at such high field strengths are not accessible experimentally \cite{Marquetand2005PCCP}. Usually, the dielectric already breaks down at field strengths on the order of 10~MV/m under normal gas-phase experimental conditions. Hence, the only way to
achieve the required field strength in an potential experiment is to use laser fields. In order to simulate these strong field effects, one has to be extremely careful. The approach described above, where the field was included in the electronic Hamiltonian, is computationally too expensive. The reason is that the field strength entering in the calculations is time dependent, and as such, it would be necessary to recalculate the complete PECs in every time step.
To circumvent this problem, the field-free potentials are usually changed \textit{a posteriori} by adding the dipole interaction (first two terms in \ref{eqn:W}). It has to be noted that one uses a Taylor expansion to describe the change of the potential energies with respect to the field strength~\cite{Atkins2003} and that higher terms like the polarizability may be needed to describe the effects of strong fields. However, it is often difficult to obtain good values especially for the polarizabilities. One reason is that tiny errors in the calculation of the energy have a dramatic effect on this second-order property. Particularly at CIs, problems can arise.
As we treat the laser interaction according to \ref{eqn:W}, we check for the quality of our curves as follows. We  diagonialize the matrix \textbf{W} for a field of $E_0=$ 10~GV/m  and compare the obtained dressed potentials (\ref{fig:1D-field}c) to the ones calculated by incorporating the interaction directly into the electronic Hamiltonian (\ref{fig:1D-field}b).

The qualitative picture of the dressed states resembles very nicely the field
perturbed PECs calculated with MOLPRO. However, we note that it is mandatory to include terms at least
up to the second term of the Taylor expansion in the electric field
interaction (the polarizability), since the dipole interaction alone is not sufficient for the regarded field strength.\\

\subsection{Quantum dynamics in the presence of an external dynamic field} 

\begin{figure*}
\centering
\includegraphics[width=15cm,keepaspectratio]{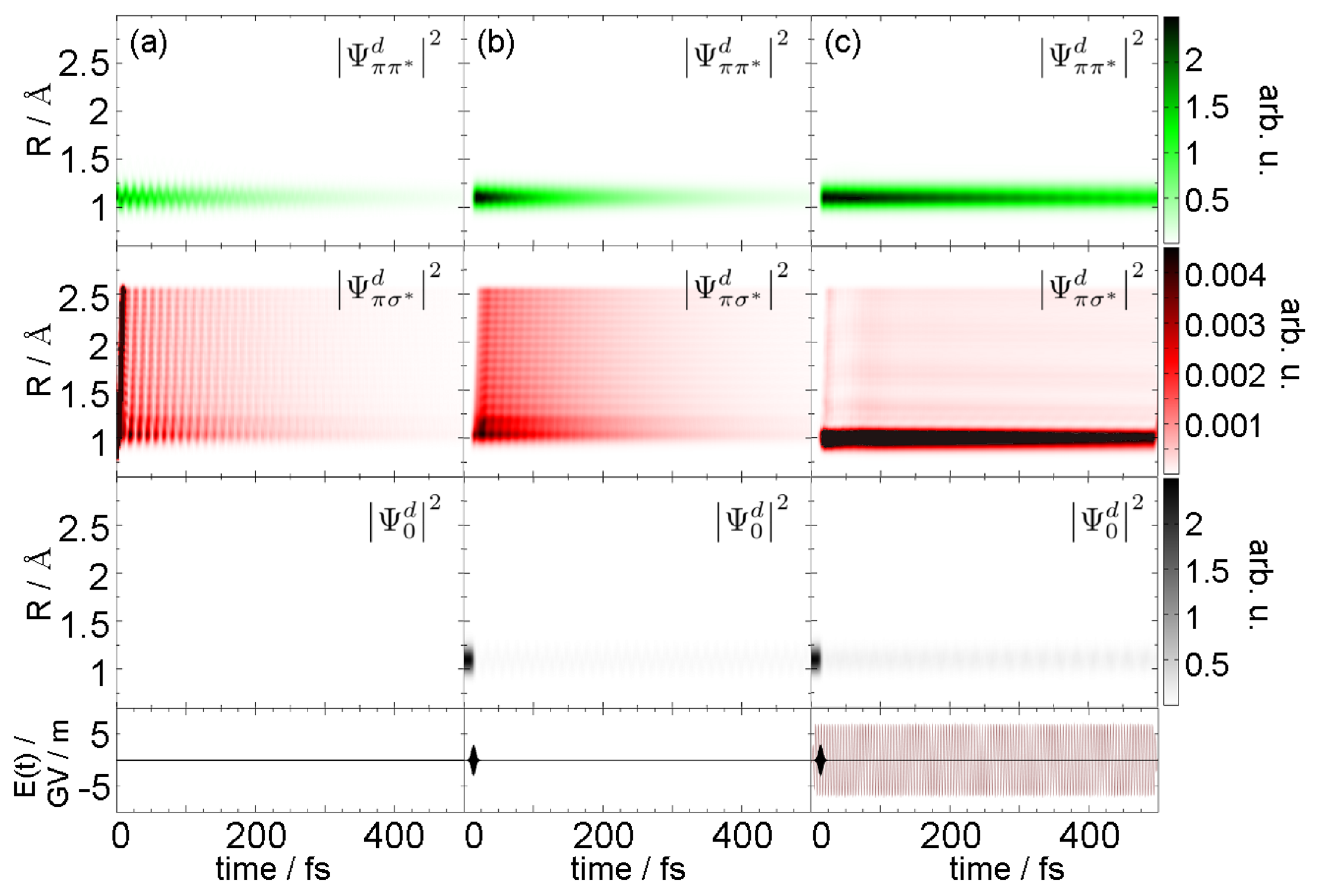}
\caption{\label{fig:wfsq} \footnotesize{Time evolution of the wavepacket in the
diabatic representation, $\left| \Psi_i^d(t) \right|^2$, in each diabatic state shown for three representative cases: (a) $\delta$-pulse only, (b) UV pulse
only, and (c) UV + control-pulse.}}
\end{figure*}

We now turn to describe the quantum dynamics influenced by a control laser. As discussed above, our goal is to prevent H-dissociation and therefore trap the molecule in a dressed state. We consider a control scheme where what we label the control laser interacts with the molecule first and only then, a UV pulse transfers population to the dressed excited states already prepared.  In order to understand the involved processes, the time-dependent probability densities $\left| \Psi_i^d(t) \right|^2$ are plotted for three scenarios in \ref{fig:wfsq}. There, we use the diabatic representation, first, because it is the natural picture for our simulation procedure as described above, and second, because the character of the nonadiabatic coupling is such that the major part of the dynamics takes place in the diabatic states. Note the different scales for the density amplitudes in \ref{fig:wfsq}. The three cases are:

Case (a). A $\delta$-pulse is used to excite population to the adiabatic state potential $V^{ad}_2$ only and no control field is present (\ref{fig:wfsq}a). This scenario also serves to compare with previous semiclassical simulations~\cite{Kinzel2011TBP} and therefore validate our 1D model.

Case (b). No control field is yet present, but a UV pulse of finite duration is employed to transfer population to all the considered excited states in a realistic fashion according to the calculated transition dipole moments (\ref{fig:wfsq}b). The employed pulse has a gaussian shape centered at $t=14$~fs with a
full-with-half-maximum (FWHM) of 7 fs, wavelength $\lambda^{UV}=$ 128~nm and a
field strength $E_0^{UV}=$ 3~GV/m.

Case (c). A control laser with a wavelength of $\lambda^{control}=$ 1200~nm is switched on from $t_i=$ 0~fs to $t_{c1}=$ 7~fs in a sinusoidal fashion and then stays at a constant field strength $E_0^{control}=$ 7~GV/m (\ref{fig:wfsq}c). The parameters for the UV pulse are the same as in case (b) except that the wavelength has been changed to $\lambda^{UV}=$ 132~nm to account for the potential energy shift induced by the control field.

\begin{figure}
\centering
\includegraphics[width=8cm,keepaspectratio]{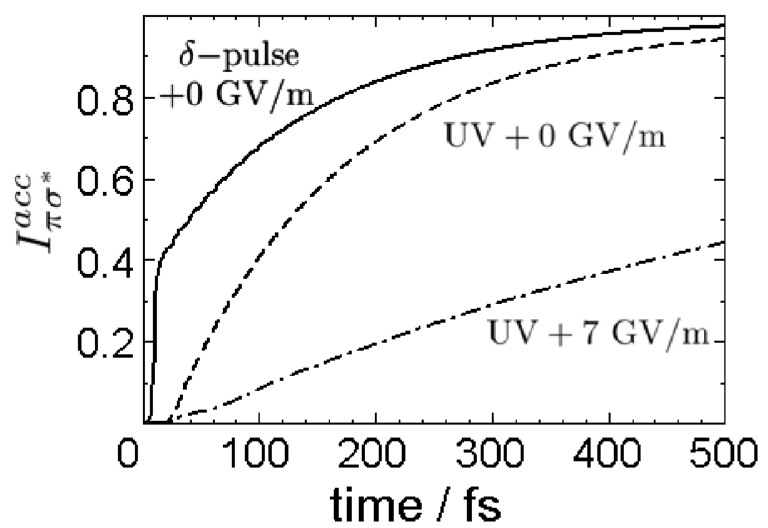}
\caption{\label{fig:flux} \footnotesize{Accumulated wavefunction flux vs. time
in the diabatic {\ps} state for case (a) $\delta$-pulse only, (b) UV pulse
only and (c) UV + control-pulse.}}
\end{figure}

The corresponding time-accumulated wavefunction flux in the diabatic {\ps}
state for each case (a), (b) and (c) according to \ref{eqn:flux} is shown in \ref{fig:flux}. This flux then
corresponds to the population that dissociated in this state. \\ 

In case (a), the vibrational ground state wavefunction is excited to the adiabatic potential $V_2^{ad}$ using a $\delta$-pulse. In the diabatic representation, this is equivalent to approximately 35\% population in the spectroscopically dark {\ps} state at time $t=$ 0~fs. A $\delta$-pulse excitation to an adiabatic state is typically employed in semiclassical simulations and the excited-state wavepacket here matches the initial conditions for the
trajectories run in Ref. \cite{Kinzel2011TBP}. As seen in \ref{fig:wfsq}a, the population in the diabatic {\ps} state is dissociating very fast within
the first 10 fs, leading to the conclusion that all molecules once transferred to this state will undergo dissociation very rapidly. This finding is not surprising recalling the repulsive
character of the {\ps} state. The rest of the wavefunction located initially on
the diabatic bright {\pp} state is oscillating in $R$ due to the displaced potential
minimum with respect to the FC geometry. During every oscillation, the {\pp}/{\ps} crossing is accessed and some portion of the wavefunction is transferred nonadiabatically to the {\ps} state leading to rapid dissociation. Within the first 50 fs already 54\% of the initial population is dissociated through the {\ps} state. This fits very nicely with the results of the trajectory simulation, which showed that 57\% of the trajectories undergo atomic hydrogen dissociation within 50 fs
\cite{Kinzel2011TBP}. In the present quantum dynamical simulation, ca. 90\% of the molecules have undergone dissociation after 200 fs, whereas at the final time of the propagation (500~fs) almost the complete population has dissociated. From \ref{fig:flux} we can also infer time constants ${\cal T}$ for the build-up of the dissociated products. We fit the corresponding curves according to $1-e^{-(t-t_0)/{\cal T}}$, where $t_0$ is the offset from $t=0$. For the process induced by the $\delta$-pulse, we see a biexponential build-up with a fast and a slow part. The corresponding time constants are obtained as ${\cal T}_{\delta\mathrm{-pulse}}^{\mathrm{fast}}= 9.36 \pm 1.98$~fs with an offset $t_0= 2.20 \pm 0.82$~fs and ${\cal T}_{\delta\mathrm{-pulse}}^{\mathrm{slow}}= 145.25 \pm 0.25$~fs with an offset $t_0= -62.75 \pm 0.29$~fs.

Using a resonant UV pulse (case (b)), 95\% of the population in the ground
state is excited to the bright {\pp} state. Although less population is initially found in the {\ps} state compared to case (a), the dissociation proceeds on a similar time scale and to a similar extent. To support this finding, we deduce a time constant also for this case from \ref{fig:flux}. The corresponding time constant is determined as ${\cal T}_{\mathrm{UV}}= 159.64 \pm 0.49$~fs with an offset $t_0= 15.61 \pm 0.30$~fs which very well agrees with ${\cal T}_{\delta\mathrm{-pulse}}^{\mathrm{slow}}$. As the degeneracy is located very close to the FC point, population is almost constantly transferred to the {\ps} state, which directly leads to dissociation. This process is effective even if the wavepacket is not moving considerably. After 50 fs 25\%, after 200 fs 70\% and at the end of the propagation time all of the excited population has dissociated.

In case (c), the control laser with a field strength $E_0^{control}=$ 7 GV/m is turned
on and as a result the potentials are strongly shifted. A net Stark shift of -0.3 eV is observed between the ground state and the bright excited state compared to the unperturbed case. Hence, the UV pulse needs to have a wavelength of $\lambda^{UV}=$ 132~nm, instead of 128~nm as in the UV-only case, to match the resonance condition. After excitation to the bright {\pp} state in the presence of the control field, the wavepacket can
indeed be trapped in the {\pp} state for much longer times than in cases (a) and (b). At the end of the propagation time, only 45\% of the total population have undergone dissociation. From the build-up of dissociated products, as depicted in \ref{fig:flux}, again a time constant can be deduced as described above. The obtained value of ${\cal T}_{\mathrm{UV+control}}= 811.33 \pm 1.19$~fs with an offset of $t_0= 21.92 \pm 0.33$~fs is much higher than the other derived constants (see above) as desired.
Yet, although the crossing is shifted to larger distances and the distorted potentials favor the desired trapping, some portion of the wavepacket can still cross to the dissociative {\ps} state and a constant loss of population is observed. However, more than 50\% of the population can be preserved in the {\pp} state during times long enough to e.g. initiate a rotation around the double bond of the molecule.\\

\textit{Fitness landscapes of reduced dimensionality}:
As we have seen above, introducing a non-resonant strong laser field with a
strength of 7~GV/m in the quantum dynamics simulations increases the lifetime of
the wavepacket in the spectroscopic {\pp} state significantly. In order to check the influence of the control field strength $E_0^{control}$, a two-dimensional pulse parameter scan is carried out. The first dimension is the field strength of the control laser scanned in the range from 0 to 20~GV/m using a step size of 1~GV/m. Since the overall Stark
effect results in shifted dressed states, where the bright {\pp} state is shifted more extensively than the others, the wavelength of the resonant UV pulse needs to be adjusted in each case. Thus, the second dimension of the scan is
$\lambda^{UV}$, which is varied from 128 to 164~nm with a step size of 2~nm.
All other parameters of the two laser pulses are kept as described above
(see ``case (c)''). 

\begin{figure*}
\centering
\includegraphics[width=15cm,keepaspectratio]{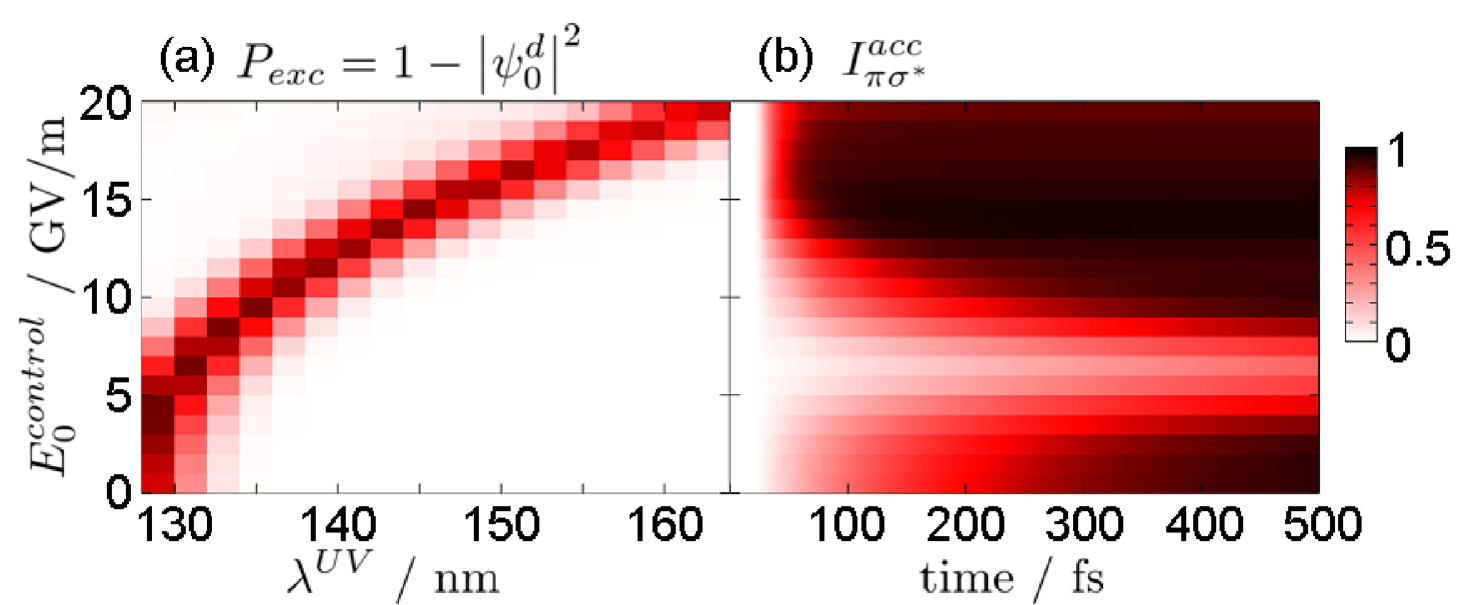}
\caption{\label{fig:maps} \footnotesize{(a) Excited population ($P_{exc}= 1-|\Psi_0^d|^2$ at $t=$ 50~fs) vs. $E^{control}$ and
$\lambda^{UV}$. (b) Accumulated flux in the {\ps} state vs. $E^{control}$ and time at
$\lambda^{UV,max}$ where $P_{exc}$ is maximum.}}
\end{figure*}

The result can be visualized in a fitness landscape of reduced dimensionality~\cite{Chakrabarti2007IRPC,Marquetand2007EPL,Marquetand2008JCP}, where we map the amount of excited-state population $P_{exc}=1-|\Psi_0^d|^2$ after 50 fs (i.e. the UV pulse is over) as a function of the wavelength of
the UV pulse $\lambda^{UV}$ and the field strength of the control
laser $E_0^{control}$, see \ref{fig:maps}a. We observe an overall shift of the
excitation energy of -2.1 eV when going from 0 to 20 GV/m, indicating a strong Stark
shift of the spectroscopic {\pp} state.

With the information about the required excitation wavelength at hand, we now take a look at the amount of population that dissociates at each particular field strength, i.e. the time-accumulated flux in the {\ps} state versus the strength of the control laser
field. Here, we consider only those cases in which the UV laser is in resonance
with the dressed {\pp} state. In \ref{fig:maps}b, the time evolution of the accumulated flux (or dissociated
population) is shown depending on the strength of the control laser,
$E_0^{control}$.

As the field strength parameter of the control laser increases, the amount of
atomic H-dissociation decreases significantly until a
field strength of $E_0^{control}=$ 7 GV/m. Hence, the wavepacket is effectively
trapped on the dressed {\pp} state, and its lifetime in this state increases.
With a field strength of 6 and 7 GV/m almost 95\% of the population are still
trapped after 50 fs, whereas 54\% are already dissociated when the control
laser is turned off. The torsional half cycle to go from one enantiomer to the
other in 4MCF has been estimated as ca. 150 fs \cite{Alfalah2010CP}.
Taking this time into account, the first 50 fs after the
excitation are sufficient for the wavepacket to decide which relaxation path to follow. Hence,
trapping the wavepacket for the first 50 fs is essential to block the undesired H-dissociation pathway.

\begin{figure}
\centering
\includegraphics[width=7cm,keepaspectratio]{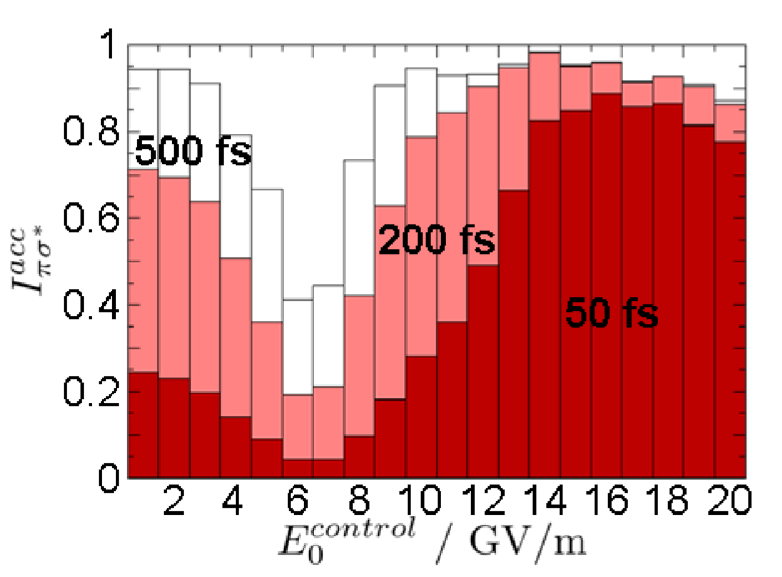}
\caption{\label{fig:bars} \footnotesize{Accumulated flux $I^{acc}_{\pi\sigma^\ast}$ vs. $E_0^{control}$ at times as indicated. The wavelength for the excitation laser is adjusted for each field strength of the control laser according to \ref{fig:maps}a.}}
\end{figure}

Interestingly, we also observe that the dissociation process can also be accelerated if the field strength of the control laser is increased above 7 GV/m. For an easier analysis, cuts through the landscape given in \ref{fig:maps}b are plotted as a bar chart in \ref{fig:bars} at times $t=$ 50~fs, $t=$ 200~fs, and $t=$ 500~fs. There, it becomes obvious that at field strengths larger than 14~GV/m the dissociation process is already almost completed after the first 50~fs.  One explanation for this behavior could be a resonant transition from the {\pp} to the {\ps} state induced by the control laser. Even if the NRDSE considers a nonresonant field, the resonance condition will almost always be met at some point along the potential energy surfaces~\cite{Marquetand2011FD,Sanz-Sanz2011FD}.
Another reason can be inferred from the Landau-Zener theory, which claims that more population is transferred if the momentum of the wavepacket at a crossing point is higher. At high field strengths, we are in the impulsive regime where the wavepacket experiences a kick due to the shifting potentials, and thus, gains a high momentum, and it is more efficiently transferred to the dissociative {\ps} state.

\section{Conclusion}
\label{sec:sum} 

In this paper we have performed quantum dynamical wavepacket propagations in the presence of external fields for (4-methylcyclohexylidene) fluoromethane (4MCF). The potentials were computed at the CASSCF level of theory just like the dipole moments and polarizabilities. The latter are important for the correct modelling of strong laser field interactions.

We have shown that laser control of 4MCF is possible by the means of the nonresonant dynamic Stark effect. 4MCF possesses two enantiomers connected by the rotation around the double bond. However, we have demonstrated in previous studies~\cite{Kroener2003PCCP,Kroener2004CP,Fujimura2004CPL,Alfalah2010CP,
Kinzel2011IJQC,Kinzel2011TBP} that after excitation several competing pathways to the torsion exist, among which H-dissociatian is the most important one~\cite{Kinzel2011TBP}. If the molecule is electronically excited to its first bright {\pp} state, where a rotation around the double bond is enabled, a conical intersection (CI) with a dark {\ps} state opens up the competitive dissociation channel. Using Stark control, the potentials can be distorted and the potential crossing can be shifted away from the Franck-Condon (FC) region in a fashion that the population is mainly trapped in the {\pp} state on time scales long enough to induce the desired torsion.
As a next step, we intend to incorporate also the torsional coordinate in our simulations in order to devise an efficient laser-induced \textit{cis}/\textit{trans} isomerization of 4MCF.


\acknowledgement

This work has been supported by the Deutsche Forschungsgemeinschaft (DFG) in the frame of a trilateral cooperation between Israel, Palestine and Germany under the project GO 1059/7-3 and
the Friedrich-Schiller-Universit\"at Jena. The authors thank Monika Leibscher and Jes\'us Gonz\'alez-V\'azquez  for interesting discussions. Generous allocation of computer time
at the Computer Center of the Friedrich-Schiller-Universit\"at is gratefully
acknowledged.


\providecommand{\latin}[1]{#1}
\makeatletter
\providecommand{\doi}
  {\begingroup\let\do\@makeother\dospecials
  \catcode`\{=1 \catcode`\}=2 \doi@aux}
\providecommand{\doi@aux}[1]{\endgroup\texttt{#1}}
\makeatother
\providecommand*\mcitethebibliography{\thebibliography}
\csname @ifundefined\endcsname{endmcitethebibliography}
  {\let\endmcitethebibliography\endthebibliography}{}

\end{document}